\documentclass[aps,pra,preprint]{revtex4}
\begin{document}
\title{Too Damned Quiet?}
\author{Adrian Kent}
\affiliation{
Perimeter Institute, 31 Caroline Street N,\\
Waterloo, Ontario, N2L 2Y5, Canada.}
\altaffiliation[Permanent address: ]{
Department of Applied Mathematics and
Theoretical Physics, University of Cambridge,
Wilberforce Road, Cambridge CB3 0WA, United Kingdom.}

\date{February 2005; minor revision April 2005}

\begin{abstract}
It is often suggested that extraterrestial life 
sufficiently advanced to be capable of interstellar 
travel or communication must be rare, since otherwise
we would have seen evidence of it
by now.  This in turn is sometimes taken as indirect evidence for
the improbability of life evolving at all in our universe. 

A couple of other possibilities seem worth considering.
One is that life capable of evidencing
itself on interstellar scales has evolved in many places but 
that evolutionary selection, {\it acting on a cosmic scale}, 
tends to extinguish species which conspicuously advertise
themselves and their habitats.  
The other is that -- whatever the true situation -- intelligent species 
might reasonably worry about the possible dangers of 
self-advertisement and hence incline towards discretion. 

These possibilities are discussed here, and some counter-arguments
and complicating factors are also considered. 
\end{abstract}

\maketitle

\section{Comments: April 2011}

This article was written in early 2005, and submitted then to
Science.  Perhaps predictably enough, it was rejected.   Since
then it has languished in my filespace, while I occasionally wondered 
whether to try to make it more publishable in a peer-reviewed science 
journal (and indeed whether that was possible).

Having recently noticed 2003 and 2005 arxiv papers by Beatriz Gato-Rivera
expressing some very similar ideas \cite{bg1,bg2}, I decided now to bite the bullet and distribute it as
is.  Sometimes, one should have the courage to 
communicate hypotheses even when they are difficult or 
impossible to test any time soon or argue rigorously for.
(As Prof Gato-Rivera did: credit to her.)  
Gato-Rivera focusses on the possibility that all advanced civilisations
would deliberately and consciously choose to remain inconspicuous,
which she calls the Undetectability Conjecture.  
While the ideas in this paper obviously can fit with this premise,   
it seems to me that the case for undetectability (such as it is)
is made a little stronger by considering the evolutionary 
mechanism for selecting inconspicuity discussed below, and
its possible applicability regardless of the conscious exercise of 
intelligent choice.  
 
To arxiv readers who feel that speculating about possible explanations for 
the lack of visible alien life goes beyond the boundaries of 
current science, I would say (a) you clearly have an arguable point, and 
part of me salutes your scientific rigour and notes that reading the 
article isn't obligatory, yet
(b) on the other hand, 
the borderline between the scientific and extra-scientific
isn't necessarily forever sharply fixed 
on this topic -- after all, the ideas {\it are} 
ultimately testable in principle by searching the cosmos and seeing
what is out there, and even now one {\it could} make  
some sort of test of the arguments' (im)plausibility, however 
inconclusive, with suitable models, and also (c) sometimes it is 
intellectually legitimate for scientists to engage in reasoning 
about subjects that go beyond the current boundaries of experimental
or observational science: we just need to be clear that's what 
we're doing when we do that (and clear too about the limitations 
of our arguments). 
 
In any case, whatever scientific status it merits, the article 
gives arguments for a point of 
view and hypotheses that I still think deserve more attention.   
So here it is.  

Apart from the addition of the references to
Gato-Rivera's work, and a disclaimer added to the acknowledgements,
the article below is unchanged from its April 2005 version. 

\section{Introduction} 

Fermi's famous question --- ``where is everybody?'' --- 
frames, in the plaintive cry of neglected hosts 
through the ages, a real intellectual puzzle.  
It {\it is}, on the face of it, surprising that 
aliens appear to display such little interest in the Earth and
its inhabitants.   
Are the scientific, artistic and technological
glories of our planet's civilisations not worthy of at least cursory
attention?
Is there really no exobiotic need --- aesthetic, intellectual,
comedic, psychotherapeutic, gastronomic --- that we terrestrials
could satisfy?   And even if extraterrestrial life isn't interested
in visiting us, why do we see no evidence of it elsewhere?  

It is difficult to discuss this topic intelligently, given that 
we know almost nothing
about the existence of bio-friendly environments
other than Earth, the existence or otherwise of
extraterrestrial life, the forms it might take,
or the ways in which it might behave. 
Most of our
ideas about extraterrestrials --- whether set out
in science fiction or (I would guess) in the 
policy documents of space exploration agencies --- 
are just figments of our psyches. \cite{cultureshock}
We do, though, have one source of conjectures that may at least
plausibly be useful: extrapolation from the Earth's evolutionary history.  

\section{Evolution on a Cosmic Scale?}

The aim here is to consider how life might evolve, or choose to behave,
when it is or could be conspicuous on a cosmic scale.  Here
``conspicuous'' means ``conspicuous to some independent lifeform on a
reasonable fraction of other cosmic habitats''.  The discussion
doesn't require assuming that life originally evolved or currently
exists only on planets, but I shall make the assumption anyway for
succinctness, using ``planet'' as shorthand for ``cosmic habitat'',
and taking ``independent lifeform'' as shorthand for ``lifeform having
an evolutionary history which begins on a different planet''.

One can consider the possible ways in which extraterrestrial life
might generally interact or avoid interaction without making any
assumption about where we fit into the picture.  If the aim of the
discussion is to give a possible explanation for why we haven't seen
any evidence of extraterrestrial life, though, we need to assume that
we are, by virtue of our senses and acquired science and technology,
reasonably competent cosmic observers, so that if cosmically
conspicuous life was prevalent, we would expect to see a fair fraction
of it.  Obviously, if we are just not equipped to look for the right
things, then our incompetence is sufficient to explain why we don't
see extraterrestrial life.  Even if that were very likely true, though
-- say, if our planet were surrounded by a dense nebula, allowing us
to observe the universe only by a very restricted frequency band ---
it would still be interesting and useful to consider how more
favourably situated and more competent extraterrestrials might
interact.  Knowing the answer might not, in this case, help explain
why we don't see extraterrestrial life, but it might be a useful guide
to our future actions (should we attempt to communicate or travel
outside the nebula?).

Note that in principle one could imagine 
one lifeform being conspicuous to another one, 
living on a different planet, without assuming that 
either is intelligent or technologically advanced.  

Now, it could be that cosmically conspicuous life, or life capable of
interstellar travel, or even just life, has evolved nowhere but on
Earth.  It could also be that, although cosmically conspicuous life
has evolved independently at many locations in the cosmos, its
evolution is relatively rare, and it generally survives for 
relatively short periods, so that creatures originating in different
cosmic habitats should expect rarely, if ever, to encounter evidence
of one another.  This has often been suggested, on the grounds that
lifeforms are unlikely to become cosmically conspicuous until
they develop radio technology, and that our own example suggests that
once intelligent lifeforms reach this level they are likely very soon
to acquire massively destructive weapons and (it is argued) will
proceed quickly to bring about their own extinction.  Another
possibility, often considered in science fiction, is that the cosmos
is teeming with intelligent species, who roam far and wide, but are
careful to ensure that we are unable to infer their existence, perhaps
because they have decided not to interfere with our development.

Though each of 
these scenarios may perhaps be
plausible, I do not want to discuss
them further here.   

The point of this paper is to introduce and consider some other
explanations.  Consider, for instance, the following hypotheses.
First, life has evolved relatively frequently in the cosmos.  Second,
interactions between species originating on different planets have not
been so uncommon.  Third, these interactions have often led to species
becoming extinct (or at least to previously conspicuous species
becoming inconspicuous, either because their numbers and/or
technological development have been greatly reduced, or because their
behaviour has been permanently altered).  Fourth, as a consequence,
evolutionary selection, operating on galactic or even larger scales,
has ensured that typical surviving species are not conspicuous to
typical observers who are based in another cosmic habitat and capable
of interstellar travel.

Perhaps the most obvious version of these hypotheses involves ``wars
of the worlds'' --- deliberately waged struggles between rival
intelligent and technologically advanced civilisations.  One could,
indeed, restrict discussion only to this case, assuming that only
intelligent species are able to travel from one planet to another.  It
is worth noting in passing, though, that the hypotheses do not
necessarily require such struggles to be the dominant selection
mechanism.  One could, for instance, imagine (though I am not sure 
whether it is really plausible) predatory species which have somehow
evolved to propagate over interstellar distances, perhaps somehow
guided by being somehow attuned to signs of distant life or
promising-looking habitats, although they have nothing that we would
recognise as advanced intelligence.

Similarly, it is important to note that cosmic inconspicuity need not 
necessarily arise as the result of a deliberate decision by an
advanced civilisation. 
Some intelligent species might indeed prudently make themselves
as inconspicuous, and their habitat as desolate-looking
and useless, as possible, either because they are concerned
about the possibility of interstellar 
predation, or perhaps even because they have observed it
in operation from afar.  
But others might perhaps 
simply lack the wit, gumption, or interest to advertise
their coming of age or to venture beyond their 
native environment --- and have rather greater chances of 
survival because of their lack of initiative.   
(No one said that life is fair.)  
And some species might avoid predation through the 
good fortune of happening to take such an unfamiliar
form that most other inhabitants of the cosmos would not
recognise them as living, or by using such exotic resources that their habitats
are not generally seen as valuable, and so surviving unnoticed.  

\section{Why Conspicuity?} 

Suppose now that there is indeed
some sort of competition
for resources on a cosmic scale and that some sort of 
evolutionary selection ensues.   Is it at all plausible that 
conspicuity should be an important criterion in this selection
process?   After all, on Earth, though many species do indeed
successfully employ camouflage and concealment, it is not generally true that
the most successful species live in habitats completely hidden from 
other competitor species.   Terrrestrial ecosystems are generally
characterised by complex symbioses and interactions amongst their
component species.    

One might imagine that, if any cosmic scale ecosystems actually exist, they 
would be similarly complex.  But there are some reasons to think that the
situation might be very different.  
If cosmic habitats can indeed largely be equated
with habitable planets, they are discrete and (on our
current best guesses) very widely separated, unlike
most terrestrial habitats.
And if inhabitated planets are very widely
separated, there may well be no easy way for an inhabitant
of planet A to identify inhabited planet B, unless
the inhabitants of B succeed in some way or other 
(perhaps inadvertently) in advertising their existence.  
One imagines that an inhabited planet,
together with the ecosystem it supports, 
constitutes a resource that would be valuable to
(some significant subset of the) species originating on other planets.
For instance, a sufficiently technologically advanced species
might find it worthwhile to use (the exobiotic analogue of)
genetic engineering to adapt the planet's entire ecosystem 
to their purposes.   
Of course, these purposes may not be entirely
compatible with the purposes of its original inhabitants.
In fact, the stakes may be still larger, 
and the competition correspondingly fiercer, than this 
suggests, if the inhabitants of planets A and B are
also potential or actual competitors for resources
elsewhere (whether other inhabited planets or some
other type of resource).  

In summary, there is a potential competition for resources, 
and a potential evolutionary mechanism for selecting
against those who tend to lose such competitions.
If cosmic habitats are widely enough separated that they
are very hard to find, by far the best strategy for 
a typical species to avoid defeat in such competitions may 
be to avoid entering them,
by being inconspicuous enough that no potential 
adversary identifies its habitat as valuable.  

\section{Complications}

But not so fast.   A little reflection suggests that, even if
conspicuity is indeed liable to provoke some sort
of predation from competitors, the selection dynamic
must be rather more complicated than the preceding
discussion allows.  In the first place, it is 
difficult to predate on a conspicuous habitat 
without making oneself conspicuous.   Even granted 
an exemplarily stealthy attack and takeover, the 
mere fact that the previously conspicuous species B
is no longer so gives a clue to observers
elsewhere that some other species A, 
with its own potentially interesting resources, 
may now be in occupation --- and hence that it may
also perhaps be worth exploring the neighbourhood
for other habitats that species A occupies.   

If species A manages to take over
while leaving in place whatever made the habitat
conspicuous in the first place, they might 
manage to make their conquest imperceptible.
But then, of course, the habitat would remain
conspicuous: this would not contribute 
to an evolutionary selection against conspicuous
habitats, and so would not support our attempt
to explain the lack of conspicuous extraterrestrials.

A really cautious predator might perhaps try to 
take over species B's habitat while giving the 
impression that species B had self-destructed.
This might or might not be believed: however 
good the cover story, it would presumably 
lose credibility if a number of independent species on different
habitats in a given region appeared to self-destruct
within a statistically implausibly short time interval.   

If B's takeover is detected or inferred by species C, 
they might be tempted to jump
in.  But so might species D, E, and so on. 
Considering this possibility might thus deter 
not only species C, but also species B. 
(Or evolution might have selected against
this behaviour.)  But then, if species B,
C and so on are all deterred, species A 
is after all left alone, happily 
broadcasting its existence to the cosmos.

Our original scenario appears to be in danger of 
self-contradiction.  
But the inconsistency arises only if one
contemplates an absolute law stating that any form
of self-advertisement is certain to 
provoke attack from competitors which
leads to extinction.  A more reasonable
picture is of a cosmos with species
which behave with varying degrees of
caginess: some predate on any 
conspicuous habitat, some attempt to 
observe predators before perhaps
predating on them, some look for
evidence of second (or third, or higher) level
predation and observe for a long time 
before deciding whether to become 
involved, and some just sit tight and
remain as inconspicuous as possible, hoping to avoid predation.  
In such a world, all species are 
playing a game with very imperfect
information.  It would be very difficult
to produce a model that convincingly
predicts the likelihoods and spatial
distributions of the various strategies,
since the answer surely depends on 
many unknowns (for example: 
what is the 
distribution in time and space   of 
the evolution of cosmically 
conspicous species?  what is 
the distribution of their 
capacities to predate and defend?  
and what is the distribution of the
strategies they are initially predisposed
to adopt?).  
The correct formulation of the hypothesis 
is simply that evolution has 
very significantly suppressed cosmic conspicuity.  

Obviously, this hypothesis may simply be wrong. 
It is not, I must admit, evident that, even if 
life did indeed originate on many different widely
separated planets, the suppression of cosmic conspicuity
would be the likeliest outcome of an evolutionary competition,
assuming one did indeed ensue.  
But the hypothesis is certainly not logically inconsistent and it 
seems to me not entirely implausible, especially in comparison
to other proposed solutions to the Fermi problem.  

A further complication, which incidentally 
lends at least slight support to the plausibility of 
the hypothesis, is that species may be 
induced to predate on conspicuous near
neighbours even if their general strategy
is to remain inconspicuous and avoid
predation.   Noisy
neighbours are liable 
to attract unwelcome attention to the 
neighbourhood.   One could perhaps 
run as far away as possible, but this requires finding
another unoccupied and inconspicuous habitat.  
Not only may this be hard, and thus dangerous in itself,
but there is the added danger that one risks becoming 
conspicuous to predators during the search.
Eliminating noisy neighbours
is, at least arguably, less dangerous than
leaving them alone: they might not yet have
been noticed by some potentially dangerous
predators, and one might presume that they
are only going to become more conspicuous,
and hence more dangerous to have around, 
unless one intervenes.  And even if 
they have already been noticed, eliminating 
them does not necessarily add to the danger:
the overall risk of more powerful predators 
arriving in the neighbourhood, and noticing you,
may not be increased.

\section{Deliberate inconspicuity} 

Intelligent species do not have to follow their 
instincts blindly.  Reflecting on 
these possibilities, and awareness of the
uncertainties involved, might deter many
intelligent species from interstellar exploration, 
and persuade them to remain as inconspicuous 
as possible on their home planet. 
A typical rational species should conclude that, 
if life in the cosmos is common, the chances
are that there are species elsewhere that are
more highly developed and likelier to prevail
in any struggle that might develop.   
If the cosmos is, in fact, a competitive and
dangerous place, a typical species 
thus ought, rationally, to be pessimistic 
about its own chance at prevailing, if it 
enters into the cosmic fray. 

The relevant probabilities are 
unknowable, but even if they are pretty
small, the cost (likely extinction) is 
so high that the possible gain of new
habitats and new knowledge may not seem
adequate compensation, unless perhaps 
one's situation is already truly desperate.  

A thoughtful species might, at the very least, decide 
to wait and observe for a long while, in 
the hope of getting some evidence about
the state of life elsewhere, before 
taking steps that would advertise its own existence.     

\section{Possible scenarios} 

Where does this all leave us?  What could the situation
be in, say, our own galaxy?   Maybe, of course, we are
the only species in the galaxy (resident or visitor) 
capable of space travel, and nothing we do in the 
foreseeable future is going to affect our conspicuity 
to any potentially dangerous species.  But the possibilities 
considered here also suggest less stable scenarios.  
We ought at least to consider them seriously and be
convinced that they really are negligible 
before neglecting them.   

One could imagine, for instance, a mixed population 
of cautious  predators and cautious stay-at-homes.  
Assuming there is currently no dominant predator,
any predators which attempted dominance in the past
must have come to grief.  (Perhaps this seems unlikely:
if it was defeated by another predator, why would that
predator not have come to dominate?  And is it really
plausible that a very powerful but reticent 
stay-at-home could, when threatened, have taken out a predator with 
galactic ambitions?)  
Advertising our existence in such an environment
would be risky: a predator species might decide it could afford
to predate on us, and even a reticent neighbour species might decide
it could not afford to leave us attracting the attention
of predators to the neighbourhood.  
 
One could also imagine a situation in which one predator
species dominates (perhaps co-existing with successfully
inconspicuous stay-at-homes) but remains as stealthy as possible. 
A rational predator species might well do just that, adopting
a predatory strategy which as far as possible eliminates 
unnecessary risk.   Advertising our existence in this
environment would most likely be suicidal.   

A third possibility is that most or all species are 
cautious stay-at-homes, fearful of interaction.  
It is unclear what result advertising one's existence
would have in such an environment, but one should
probably not expect a welcoming response.   

\section{Some Counter-arguments}

\subsection{Aliens are civilised} 

One common counter-argument arises from scepticism that species will wish to 
carry on competing by the time they have become sufficiently advanced
to attempt interstellar travel.  
Would the wisdom of co-operation 
not be clear,
and would respect for exobiotic cultures not be a universally
accepted value, by that point in a species' evolution?  
Could any advanced species really destroy the unique and irreplaceable 
culture of another, in the process gravely altering an ecosystem
that was millions or billions of years in the making? 
Morality aside, would enlightened self-interest not 
militate against such vandalistic aggression?  

These are nice and optimistic presumptions, and one would like to
believe they are right, but the evidence of the last few millennia on
Earth does rather give pause for thought.  Perhaps there really are
cosmic civilisations worthy of the name.  But everything alive in the
cosmos has, presumably, evolved to compete for resources in
environments where they are not always plentiful, to reproduce
successfully, and to give its offspring the best chances in life.
Almost every aspect of the psychological makeup of aliens may be
utterly unrecognisable to us: indeed the term itself may not apply in
any recognisable sense.  But the best guess we can make, extrapolating
from life on Earth, is that any species which controls significant
resources is likely to have maintained a powerful urge to live and to
multiply.

Perhaps, nonetheless, enlightened self-interest prevails.
Perhaps co-operation is, and is almost universally recognised
as, a better survival strategy than 
predation; perhaps a combination of consensus and 
enforcement has brought about some sort of modus vivendi.   
If there is intelligent life out there then we 
had better hope so, and hope also that
other civilisations are tolerant of
our naivet\'e.   It doesn't seem a good advertisement for 
our value as recruits to cosmic society that we have not
yet even seriously contemplated the alternative scenarios.   

\subsection{Self-advertisement signals peaceful intention} 

It is sometimes suggested, when our own attempts to 
contact extraterrestrials are discussed, that the 
sort of trusting openness we have displayed to date
is the best policy.  Since we 
must realise that aliens are likely to be more 
advanced than us, we anyway
could not hope to prevail
in competition with them. 
By displaying our trust, we show that our own 
intentions are peaceful and our desire is to learn 
and collaborate, and hence that they have nothing to fear 
from us either now or at any point in the future, 
and that we deserve their help and respect.
Should we be persuaded by this argument?  More to the
point, might another species who receives our signals be?  

I rather doubt it, unless extraterrestrial civilisations share our 
capacity for spasmodic outbreaks of truly epic self-delusion. 
In the first place, as our own  
history sadly illustrates very well, aggressors 
may not care much whether a weaker group (or species) 
has peaceful intentions or not: they tend rather to latch on to 
the point that their relative strength allows them to 
capture and exploit its resources regardless. 
Secondly, our superficially trusting openness
does not actually credibly signal that our intentions
are necessarily peaceful.   The most cursory inspection of 
present day Earth would show the discerning alien
that our intentions are, at best, mixed and fickle.  
If we happened to 
come upon a weaker species inhabiting a resource-rich
planet similar to Earth somewhere in our vicinity, our leaders would
no doubt 
make the right noises about our peaceful 
and noble intentions, and a sizeable number
of us would sincerely agree --- but history
suggests it would be rash to bet against 
our colonising and exploiting the planet 
in the long run.   

Our openness to date in advertising ourselves to 
hypothetical extraterrestrials 
seems to me to demonstrate only our lack of thought and
self-knowledge, not to mention a
decidedly tactless lack of consideration for our 
cosmic neighbours who, if they exist,
may not share our insouciance about attracting
possible predators to the region.   

\subsection{Interstellar conquest isn't worth it}

Another common counter-argument is that it hardly matters whether
aliens are malevolent or benevolent, since any species which 
is technologically advanced enough to constitute a threat to us will  
also surely --- since technology develops quickly, and the chances
of two independent species being at similar levels of development
is very small --- be so advanced that it has no use for the pitiful
resources we and our fellow terrestrials have to offer.   
bossibly --- but again, terrestrial history gives one pause
for thought.   Great empires do take an interest even in 
rather small and unprepossessing territories.  They can
do so, often, precisely because they are techologically
advanced: conquest thus requires few personnel and small
resources, relative
to their capacity.   It may indeed not be worthwhile 
for the hyper-advanced Sirians to send a large fraction of their 
interstellar battle fleet to conquer Earth and rearrange its 
ecosystem into a productive Sirian-friendly habitat.  
But might the rewards not justify the effort for 
three bored undergraduate Sirians armed with borrowed
lab equipment and looking for a mildly remunerative
holiday project?    

\section{Conclusion}

The Fermi problem is a real puzzle, and it seems to me 
that none of the proposed solutions (including the one
suggested here) is so compelling as to inspire confidence
that it is clearly right.   Yet there must be 
a solution, of course.   The relevant question, thus, it seems to me, 
when considering any proposed solution is not 
whether it is completely convincing, but whether
there is another solution which is clearly more
compelling.  In other words, does the proposed solution
belong on the list of contenders?  If it does, then 
we had better take it seriously as a possibility when
considering policy.   

We do not seem, as a species, to be particularly good at 
thinking realistically either about small risks of large
disasters or about our situation in the cosmos. 
These are seen as not quite respectable topics in the
academic world.  Almost no one is employed to worry about them,
and the prevailing mood of postmodern irony makes it 
hard for us to take them quite seriously.  
But they deserve to be taken seriously nonetheless. 
   
We have, famously, taken a few steps actively to advertise our existence:
the 1974 Arecibo message \cite{arecibo} 
beamed towards the Hercules globular cluster, 
the golden records \cite{voyager} 
and plaques \cite{pioneer} 
carried aboard the Voyager and Pioneer spacecraft, 
carrying information about us and star maps directing any 
interested aliens to our solar system, and, more recently, 
the Team Encounter ``Cosmic Call'' messages \cite{cosmiccall}
sent to a number of nearby ($30-75$ light years away) stars. 

What was the point of these efforts?  Were they well justified?   
Distinguished scientists and others have argued on several occasions that,
on the contrary, they were dangerously, and potentially 
suicidally, irresponsible \cite{critics}. 
These arguments appear to have largely fallen on deaf ears, 
but they deserve repeating, and perhaps can even be amplified a 
little in the light of the above discussion.   
 
One can summarise the essential point simply enough. 
If there are no aliens out there, any efforts at
communication were obviously wasted.   Thus we 
can assume for the sake of discussion that there are aliens 
out there likely to receive the messages at some point.
The relevant parameter, then, is not the probability 
of our messages being received by aliens who might potentially
do us harm: it is the conditional probability of 
the aliens who receive the messages doing us harm, given
that the messages are indeed received (and understood to
be messages).   Can we really say that {\it this} probability 
is so negligible, bearing in mind that any such aliens appear to 
have made no reciprocal attempts to advertise their existence? 
The arguments considered above suggest that 
we cannot.  
 
The Pioneer and Voyager spacecraft are travelling
sufficiently slowly that they will pose essentially no additional risk
for a very long while.   We could, though, take steps to enforce a ban
on any further attempts along the lines of the Arecibo and Team Encounter
messages.  We might also want to take steps to intercept and 
return the Pioneer and Voyager craft some time in the next 
few millennia.  These might, at this point, only be gestures. 
But if we believe in the merits of gestures at all, they seem more 
worthwhile than most.  They would, at least, be small signals 
that we have begun contemplating the possible realities 
of our cosmic existence, and might just possibly 
be coming to realise that it would be wise to tread carefully
and politic to take care not to jeopardise the interests of our
neighbours.  If anyone happens already to be listening,
that really would be a message worth sending.   

One final point: it often seems to be implicitly 
assumed, and sometimes is explicitly argued, 
that colonising or otherwise exploiting the resources of 
other planets and other solar systems will solve our problems
when the Earth's resources can no longer sustain our consumption. 
It might perhaps be worth contemplating more seriously the possibility
that there may be limits to the territory we can safely colonise
and to the resources we can safely exploit, and to consider 
whether and how it might be possible to evolve towards a way of living 
that can be sustained (almost) indefinitely on the resources of
(say) our solar system alone.  

\vskip5pt \leftline{\bf Acknowledgments}
I would like to thank Rob Spekkens for a particularly helpful
discussion, while stressing that 
responsibility for the ideas expressed here is of course mine
alone.



\vskip10pt
\begin{thebibliography}{99}
\bibitem{bg1}
B. Gato-Rivera, arXiv:physics/0308078.
\bibitem{bg2}
B. Gato-Rivera, arXiv:physics/0512062v3. 
\bibitem{cultureshock}
For example, it seems to me that even
the most thoughtful
science fiction tends to skate over the likely social
realities of alien contact.  An obvious point which
is most often neglected is that encountering intelligent 
extraterrestrial life, {\it even if
it were benign}, would likely 
be enormously traumatic for us all.  
What would we do, and how would
our societies reorganise, on encountering a civilisation 
that may well be far more competent
than us at essentially {\it everything} we value?
\bibitem{arecibo}
See e.g. http://en.wikipedia.org/wiki/Arecibo\_message
\bibitem{voyager}
See e.g. http://en.wikipedia.org/wiki/Voyager\_Golden\_Record
\bibitem{pioneer} 
See e.g. http://en.wikipedia.org/wiki/Pioneer\_plaque

\bibitem{cosmiccall}
See e.g. http://en.wikipedia.org/wiki/Communication\_with\linebreak
\_Extraterrestrial\_Intelligence\#Cosmic\_Call\_Messages
and http://www.matessa.org/~mike/dutil-dumas.html
\bibitem{critics}
The British astronomer and Nobel laureate 
Martin Ryle, protested forcefully against
the Arecibo message, and lobbied the International 
Astronomical Union (of which he was a past president) to pass a resolution 
in condemnation.  
The distinguished Australian palaeontologist and evolutionary
biologist Michael Archer has advanced similar arguments and
warnings. 
The American diplomat Michael Michaud also
queried the wisdom of sending the Arecibo message, arguing 
(correctly in my opinion) 
that the message was fundamentally a political act rather than
a scientific experiment, and required political consideration.   

Potted summaries of these objections, and of some responses, 
can be found, for example, at 
http://www.planetary.org/html/UPDATES/seti/Contact\linebreak
/RespondingToAliens.html, 
and in 
{\it Sharing the Universe: Perspectives on Extraterrestrial Life}, 
S. Shostak and F. Drake, (Berkeley Hills Books, 1998).

In qualified support one might also cite Jared Diamond's 
discussions in the article {\it Alone in a Crowded Universe} in {\it 
Extraterrestrials: Where Are They?} (B. Zuckerman and M. Hart,
eds, Cambridge University Press, (2nd edition, 1995)
and in {\it The Rise and Fall of the Third Chimpanzee: Evolution
and Human Life} (Vintage, 2004). 
Diamond agrees that communicating with intelligent aliens would indeed be 
highly dangerous, but suggests that the point is probably moot,
as we are quite likely
the only examples of intelligent life in the cosmos.    
\end{thebibliography}
\end{document}